\def \cm{~\rm{cm}}
\def \s{~\rm{s}}
\def \km{~\rm{km}}

\def \K{~\rm{K}}

\def \erg{~\rm{erg}}

\def \yr{~\rm{yr}}

\def \kpc{~\rm{kpc}}

\documentclass[12pt,preprint]{aastex}
\usepackage{natbib}

\shortauthors{Soker}

\begin{document}

\title{A MODERATE COOLING FLOW PHASE AT GALAXY FORMATION}

\author{Noam Soker\altaffilmark{1}}

\altaffiltext{1}{Dept. of Physics, Technion, Haifa 32000, Israel;
soker@physics.technion.ac.il}

\begin{abstract}
I study the possibility that a cooling flow (CF) exists at the main phase
of super massive black hole (SMBH) growth during galaxy formation.
To ensure that jets launched by the SMBH efficiently expel gas from the galaxy,
as is required by recent results, the gas should be in the hot phase, rather than
in cold clouds.
The short radiative cooling time of the hot gas leads to the formation of a CF,
but heating by the active galactic nucleus (AGN) prevents catastrophic cooling.
Cold blobs that start as instabilities in the hot phase feed the SMBH from an extended region,
form an accretion disk, and lead to the formation of jets.
These jets can expel large quantities of gas out of the galaxy.
This cycle, that is termed cold feedback mechanism in CFs in clusters of galaxies, might
explain the correlation of SMBH to bulge masses.
Stars are formed, but at a lower rate than what is expected when heating is not included.
Such a CF is termed a moderate CF.
\end{abstract}

\section{INTRODUCTION}
\label{sec:intro}

The tight correlation between the supermassive black hole (SMBH) mass, $M_{\rm BH}$,
and the velocity dispersion, $\sigma$, of the hot component of the host galaxy
(e.g., Merritt \& Ferrarese 2001; Gebhardt et al. 2000; G\"ultekin et al. 2009 )
is well established, and so is the correlation between the SMBH mass and the
bulge mass, $M_{\rm bulge}$ (e.g., Kormendy \& Richstone 1995; Laor 2001).
It is possible that the $M_{\rm BH}-M_{\rm bulge}$ correlation is determined by
a feedback process of the active galactic nucleus (AGN) of the host bulge,
powered by the accreting SMBH.
The feedback mechanism where AGN jets (outflow; wind) suppress gas from cooling to
low temperatures and from forming stars was discussed for both cooling flows (CFs)
in galaxies and clusters of galaxies (e.g., Binney \& Tabor 1995;
Nulsen \& Fabian 2000; Reynolds et al. 2002; Omma \& Binney 2004; Soker \& Pizzolato 2005),
and in galaxy formation (e.g., Silk \& Rees 1998; Fabian 1999; King 2003;
Croton et al. 2006; Bower et al. 2008; Shabala \& Alexander 2009).
By galaxy formation I will refer also to bulge formation in a spiral galaxy,
and vice versa.

There are other attempts to explain this correlation.
For example, in their model of bulge formation by coalescence of giant
($M \sim 10^8 M_\odot$) clumps, Elmegreen et al. (2008b) considered feedback by
supernovae (SNe), but not by AGN activity.
In their model, the SMBH at the center of the galaxy is formed from the merger of
the intermediate mass BHs that reside in each clump (Elmegreen et al. 2008a).
The ratio of SMBH to bulge mass $M_{\rm BH} /  M_{\rm bulge}  \sim 10^{-3}$ originates in the
massive clumps.
In the present paper I assume that the SMBH forms by accreting mass from the ISM,
and that the $M_{\rm BH}-M_{\rm bulge}$ correlation is determined by an AGN feedback mechanism.
It is determined in the sense that the AGN activity limits the ISM mass that is eventually
converted to stars by expelling it out of the galaxy.
The ejection of large quantities of gas out of the galaxy during its formation
was found to be a necessary process in recent studies (e.g., Bower et al. 2008).

In a previous paper I tried to account for the SMBH$-$bulge masses correlation
with a feedback mechanism based on jets launched by the SMBH.
This feedback is based on narrow jets that are launched by the central SMBH, and
expel large amounts of mass to large distances.
The condition for an efficient expelling process is that the jets do not
penetrate through the inflowing gas, such that they can deposit their energy
in the inner region where the bulge is formed.
A relation between the mass accreted by the SMBH and the mass that is not
expelled, and is assumed to form the bulge, was derived (Soker 2009;
{{{  this derivation is repeated in section 4.3). }}}
It was noted that the same mechanism could operate in suppressing star formation
in CF clusters, making a tight connection between the feedback
in galaxy formation and CF clusters.
In the present paper I extend the comparison between the feedback mechanism operating
in CF clusters and during galaxy formation.
My fundamental assumption is that the correlation between the SMBH mass and the
host galaxy bulge mass is determined by an AGN feedback process that expel large
quantities of gas out from the galaxy.

In section \ref{sec:bondi} and \ref{sec:cold} I argue that for the feedback
mechanism to be efficient, as required from my fundamental assumption, most
of the gas should pass through the hot phase, unlike the case in the model studied by
Binney (2004, 2005).
In section \ref{sec:cooling} I discuss a model where the feedback between the
SMBH and star formation in the bulge occurs through a process similar to that
operating in CFs in clusters of galaxies.
A comparison of CF in clusters to the process at galaxy formation was done in the past
(e.g., Croton et al. 2006).
In the model discussed by Binney (2004; 2005), for example, heating by the AGN activity
and the ejection of large quantities of gas out from the galaxy are crucial
processes, as they are in the present model.
However, Binney (2004) proposed a model where the gas that forms star has never
been heated to the virial temperature, while that gas that was heated to the virial
temperature has negligible contribution to star formation.
In the model proposed in section \ref{subsec:cooling2}, on the other hand, large
quantities of gas in the hot phase cool to form stars.
The advantage in the presence of large amount of mass in the hot phase is that
the hot phase is more susceptible to AGN heating
(Binney 2004; Dekel \& Birnboim 2006; Cattaneo et al. 2006; Hopkins \& Elvis 2009).
I summarize the proposed model in section \ref{sec:summary}.

\section{PROBLEMS WITH FEEDBACK DRIVEN BY BONDI ACCRETION}
\label{sec:bondi}

I define the Bondi accretion radius by
\begin{equation}
R_B=\frac {2GM_{\rm BH}}{C_s^2}=
\frac{6}{5} \frac {GM_{\rm BH } \mu m_H}{kT}
\label{eq:bondi1}
\end{equation}
where $M_{\rm BH}$ is the SMBH mass , $C_s=(5kT /3 \mu m_H)^{1/2}$ is the ISM
sound sped, and all other symbols have their usual meaning.
The Bondi accretion rate for an adiabatic index of $\gamma = 5/3$ is (Bondi 1952)
\begin{equation}
\dot M_{\rm Bondi} = 0.25 \pi R_B^2 \rho C_S,
\label{eq:bondi2}
\end{equation}
where $\rho$ is the density at a large distance from the accreting body.
The characteristic inflow time is
\begin{equation}
\tau_{\rm flow} \equiv \frac{R_B}{C_s}.
\label{eq:bondi3}
\end{equation}
{{{  (The inflow velocity at $R_B$ is $\ll C_s$, and the practical inflow
time scale is longer than $\tau_{\rm flow}$.)  }}}

A condition for the existence of a Bondi type accretion flow is that the
radiative cooling time is longer than the inflow time $\tau_{\rm cool} > \tau_{\rm flow}$.
Otherwise the ISM rapidly cools, the value of $C_s$ rapidly decreases,
and the assumptions that lead to the Bondi accretion flow in a hot ISM break down.
Namely, we are in the regime of accreting cold gas, the problems of which are
discussed in section \ref{sec:cold}.
This condition reads
\begin{equation}
\frac{3}{2} \frac{nkT}{\Lambda n_e n_p} >   \frac{R_B}{C_s},
\label{eq:cool1}
\end{equation}
where $n$, $n_e$, and $n_p$ are the total, electron, and proton number density, respectively.
Condition (\ref{eq:cool1}) can be cast in the form
\begin{equation}
5.2 \frac{\mu m_H kT}{\Lambda } R_B C_s^2  \ga 0.25 \pi {R_B^2} \rho {C_s}.
\label{eq:cool2}
\end{equation}

For the cooling function $\Lambda$ I assume a zero metallicity composition
and take the temperature range $T \ga 10^7 \K$. This gives
$\Lambda_0(T) \simeq 2 \times 10^{-23} (T/10^7 \K)^{1/2} \erg \cm^3 \s^{-1}$
(Sutherland  \& Dopita  1993).
For lower temperatures and higher metallicity the conclusions will be stronger even,
as cooling time is shorter.
Substituting equation (\ref{eq:cool2}) into equation (\ref{eq:bondi2}) with the above
approximation for $\Lambda$,  gives
\begin{equation}
\dot M_{\rm Bondi}< 1.6 \times 10^{-9}     
\left( \frac{T}{10^7 \K} \right)^{1/2}
\frac{M_{\rm BH}}{M_\odot}
M_\odot \yr^{-1} .
\label{eq:cool3}
\end{equation}
The actual limit is lower even for three reasons:
(1) Some fraction of metals will be presence, increasing the cooling rate.
(2) The inflow time should be shorter than the cooling time not only from the
Bondi radius, but also from a distance of several Bondi radii.
(3) In galaxies that do not sit at the center of groups and clusters, the virial temperature
of the gas is $<10^7 \K$. For a temperature of $T=2 \times 10^6 \K$ and zero metallicity,
the cooling function is lower by a factor of 1.6, and the cooling time shorter
by a factor of $5/1.6 =3$. The numerical coefficient in equation  (\ref{eq:cool3}) is
$\sim 5 \times 10^{-10} M_\odot \yr^{-1}$.

We can compare the limit from the Bondi accretion as given in equation (\ref{eq:cool3})
with the Eddington limit
\begin{equation}
\dot M_{\rm Edd} = 2.5 \times 10^{-8}
\left( \frac{\epsilon }{0.1} \right)^{-1}
\frac{M_{\rm BH}}{M_\odot}
M_\odot \yr^{-1} ,
\label{eq:eddington1}
\end{equation}
where $\epsilon$ is the efficiency of converting mass to radiation in the accretion process.
We find that $\dot M_{\rm Bondi} < 0.1 \dot M_{\rm Edd}$.
It seems there is no time for the BH to grow if it is limited by the Bondi accretion,
as the e-folding time is
\begin{equation}
\tau_f = \frac {M_{\rm BH}}{\dot M_{\rm Bondi}} >
6 \times 10^8
\left( \frac{T}{10^7 \K} \right)^{-1/2}
\yr .
\label{eq:tauf1}
\end{equation}
To grow by a factor of 1000, the minimum time required  is $\sim 4$~Gyr, with a more likely
value of $> 5$~Gyr.
Clearly, this process cannot explain the rapid growth of SMBH during galaxy formation.

There are two possible solution to the feeding problem.
\newline
(1) Most of the ISM mass is in a cold $T \la 10^4 \K \ll T_{\rm virial}$ phase,
for which the Bondi accretion radius is much higher.
The problem with cold clouds is that they are very dense.
It is very hard to remove such clouds from the galaxy.
More over, cold clouds are more likely to form stars.
This solution makes it very hard to explain the termination of star formation.
This is the subject of the next section.
In any case, models that include Bondi accretion consider the hot $T \ga 10^6 \K$ phase.
\newline
(2) Most of the mass is in the hot phase with $T \simeq T_{\rm virial}$.
However, the SMBH is fead by cold clumps that are embedded in the hot ISM.
The clumps as a group contain a small fraction of the total ISM mass.
This solution to maintain feedback heating in cluster CFs is termed
the {\it cold feedback mechanism} (Pizzolato \& Soker 2005, 2010; Soker 2006, 2008b).
In section \ref{sec:cooling} I compare this mechanism as it operates in galaxy formation
at high redshifts to its role in low redshift cluster CFs.

\section{PROBLEMS IN MAINTAINING FEEDBACK WITH MASSIVE COLD ISM}
\label{sec:cold}

It is possible that most of the ISM mass resides in cold {{{  ($\sim 10^4 \K$) }}} clouds,
rather than being close to the virial temperature and in a hydrostatic equilibrium.
The problem in this case is that the AGN feedback efficiency is too low, because it is
extremely difficult to expel dense clouds from the inner regions of the galaxy.
In a recent attempt Hopkins \& Elvis (2009) note that AGN feedback works more efficiently
on the hot phase, and consider a chain of processes to overcome this problem.
In their model the clouds contain $\sim 90 \%$ of the mass, while $\sim 10 \%$ of the ISM
mass is in the hot phase at hydrostatic equilibrium.
A shock propagating through the hot phase causes the clouds to expand, such that
radiation pressure from the AGN is more efficient in expelling the bloated clouds.
Even in this more efficient scenario, the mass expelled in their model is not
sufficient for the feedback mechanism studied here.
Hopkins \& Elvis (2009) take the bulge mass to be $10^{11} M_\odot$, while the ISM mass
is only $10^{10} M_\odot$.
After a time of $t= 10^8 \yr$ from the start of their calculation,
the mass expelled in their model is $\Delta M_e = 5 \times 10^9 M_\odot$.
During the same period, for an efficiency of $10\%$, the SMBH mass has grown by
$\Delta M_{\rm SMBH} \simeq  10^8 M_\odot$.
This ratio of $\Delta M_e/\Delta M_{\rm SMBH} \simeq 50$, is lower by more than an order of magnitude
from the one required in feedback models that account for the bulge to SMBH masses ratio,
if the feedback is to determine the final mass of the bulge.
Had they include radiative cooling, the efficiency of expelling the ISM would be lower even.

The conclusion is that for an AGN feedback to work, most of the gas in the inner regions,
$r \la 10 \kpc \sim 0.1 R_{\rm v}$, must reside in the hot phase; $R_{\rm v}$ is the virial radius.

Although the AGN feedback requires the inner regions to be mostly
in the hot phase, the gas feeding the galaxy at larger regions can be cold.
In the cosmological cold streams model (Dekel et al. 2009a, b and references therein)
three cold, low entropy, streams of gas penetrate the virial radius without being shocked,
and reach the central region.
Shocks near the center are not resolved in the numerical simulations presented
by Dekel et al. (2009a), as their resolution is $1.4 \kpc$.
Eventually, the streams must encounter a shock wave near the center.
The typical influx rate considered by Dekel et al. (2009a) is
$\dot m \sim 30 M_\odot \yr^{-1} {\rm rad}^2$.
At a distance $r$ from the center the post-shock total (protons, electrons and nuclei) number density is
\begin{equation}
n \simeq  1
\left(   \frac{\dot m } {30 M_\odot \yr^{-1} {\rm rad}^2} \right)
\left( \frac {v}{300 \km \s^{-1}} \right)^{-1}
\left( \frac {r}{5 \kpc} \right)^{-2}  \cm^{-3},
\label{eq:den1}
\end{equation}
where $v$ is the preshock inflow velocity, scaled with the inflow velocity at
$ r \sim 5-10  \kpc$ from the center according to Dekel et al. (2009a).
The postshock temperature is $\sim 10^6 \K$.

For a temperature of $\sim 10^6 \K$, and zero metallicity, the cooling function is
$\Lambda_0 \simeq 10^{-23} \erg \cm^3 \s^{-1}$
(Sutherland  \& Dopita  1993).
The cooling time of the post-shock gas is
\begin{equation}
\tau_{cs} \simeq 3 \times 10^6
\left( \frac{\dot m } {30 M_\odot \yr^{-1} {\rm rad}^2} \right)^{-1}
\left( \frac {v}{300 \km \s^{-1}} \right)^{3/2}
\left( \frac {r}{5 \kpc} \right)^{-2}
\left( \frac {\Lambda_0}{10^{-23} \erg \cm^3 \s^{-1} } \right)  \yr,
\label{eq:time1}
\end{equation}
where the dependence of the post shock temperature on velocity has been used.
The half opening angle of a stream is $\sim 10-15 ^\circ$; I scale with $12.5 ^\circ$.
The pressure of the post-shock gas is larger than that of its surroundings,
and it expands to the sides in a typical time of
\begin{equation}
\tau_{f} \simeq  \frac{r \sin 12.5^\circ }{v} = 3.5 \times 10^6
\left( \frac {v}{300 \km \s^{-1}} \right)^{-1}
\left( \frac {r}{5 \kpc} \right)   \yr.
\label{eq:time2}
\end{equation}

At $r \simeq 5 \kpc$ the radiative cooling time is about equal to the expansion time of the
postshock region $ \tau_{cs} \simeq \tau_{f}$.
Heating by the AGN will make the formation of a shock wave more likely.
{{{  Cantalupo (2010) showed that ionization by star forming regions can remove
cooling agents (ions) from the gas and by that efficiently reduces radiative cooling and
transforms a cold mode accretion into a hot one.
The effect is larger even when radiation from the AGN is considered.
The effect of removing cooling ions comes in addition to the heating discussed next. }}}
Let the gas mass in the inner region be about equal to the stellar mass $\sim 10^{11} M_\odot$.
With a total number density of $ n \simeq 1 \cm^{-3}$ the mass resides within a radius of $\sim 10 \kpc$.
To maintain this mass that has a radiative cooling time of $\sim 3 \times 10^6 \yr$
 at a temperature of $ T\simeq 10^6 \K$, requires a heating power of
$\dot E_{\rm H} \simeq 4 \times 10^{44} \erg \s^{-1}$.
With an efficiency of $ 10 \%$ in converting mass to energy, the accretion rate onto the SMBH is
$ \sim 0.1 M_\odot \yr^{-1}$. In $ 10^9 \yr$ the SMBH gains a mass of $\sim 10^8 M_\odot$.
In reality, the amount of ISM mass will be lower, as part of it is continuously forming stars,
while some of the incoming gas is expelled back to large distances.
This crude calculation
{{{  and the reduction in radiative cooling rate as discussed by Cantalupo (2010) }}}
show that heating by the AGN can maintain a shock wave at a radius of $r \ga 5 \kpc$.

There is no problem in the scenario proposed here that the streams penetrate the virial radius,
at $ R_v \simeq 50-100 \kpc$, as in the simulations presented by Dekel et al. (2009a).
However, the model here does require that the streams eventually
are shocked at $r \ga 0.1 R_V$, and a hot pseudo static atmosphere is formed.
The presence of hot static atmosphere around an already grown SMBH was assumed before,
e.g., Short \& Thomas (2009).

\section{A COOLING FLOW PHASE}
\label{sec:cooling}

\subsection{The cold feedback mechanism in low redshift clusters}
\label{subsec:cooling1}

In the cold-feedback model for clusters of galaxies (Pizzolato \& Soker 2005, 2010; Soker 2006; 2008a)
mass accreted by the central SMBH originates in non-linear over-dense blobs of gas
residing in an extended region of $r \sim 5-50 ~{\rm kpc}$; these blobs are
originally hot, but then cool faster than their environment and sink toward the center
(see also Revaz et al. 2008).
The mass accretion rate by the central black hole is determined by the cooling time
of the ICM, the entropy profile, and the presence of inhomogeneities (Soker 2006).
Most important, the ICM entropy profile must be shallow for the
blobs to reach the center as cold blobs.
Wilman et al. (2008) suggest that the behavior and properties of the cold clumps they
observe in the cluster A1664 support the cold feedback mechanism.

The cold feedback mechanism in clusters has the following consequences.
\newline
{(1)} Cooling flows do exist, but at moderate mass cooling rates:
            the \emph{moderate CF model}  (Soker et al. 2001).
Indeed, in many CF clusters the heating cannot completely offset cooling
(e.g.,  Clarke et al. 2004; Hicks \& Mushotzky 2005; Bregman et al. 2006; Salome et al. 2008;
Hudson et al. 2010) and some gas cools to low temperatures and flows
inward (e.g., Peterson \& Fabian 2006) and forms stars (Wise et al. 2004; McNamara et al. 2004).
Star formation is prominent when the radiative cooling time of the hot gas is short
(Rafferty et al. 2008).
These observations suggest that indeed cooling of gas from the hot phase
to low temperatures does take place, including star formation.
{{{  It is important to note that star formation and the feeding of the SMBH occur simultaneously. }}}
\newline
{(2)} The cold feedback mechanism explains why real clusters depart from
   an `ideal' feedback loop that is 100\% efficient in suppressing cooling and star formation.
   Simply, the feedback requires that non-negligible quantities of mass cool to low temperatures.
   Part of the mass falls to small radii. Part of this mass forms star, another part is ejected back,
    and a small fraction is accreted by the central SMBH.
\newline
{(3)} Part (likely most) of the inflowing cold gas is ejected back from the very inner region.
   This is done by the original jets blown by the SMBH (Soker 2008b). The ejection of this
   gas is done in a slow massive wide (SMW) bipolar outflow, which are actually two jets.
 {{{  The basic mechanism is that the jet does not puncture a hole in the ICM, but rather
   deposits its energy in the inner region. Wide enough jets deposit their energy in the inner
   regions rather than puncturing a hole and expanding to large distances.
   Rapidly precessing jets or a relative motion of the medium also prevent such a puncturing.
    In the case of the formation of low density bubbles in the ICM, it has been shown that
    in addition to SMW jets, precessing jets (Sternberg \& Soker 2008a; Falceta-Goncalves et al. 2010)
    and a relative motion of the medium (Morsony et al. 2010) also lead to the inflation of bubbles
     close to the center.  }}}
   The most striking example of a SMW outflow is presented in the seminal work of Moe et al.
(2009; also Dunn et al. 2009).
By conducting a thorough analysis, Moe et al. (2009) find the outflow
from the quasar SDSS J0838+2955 to have a velocity of $\sim 5000 \km \s^{-1}$,
and a mass outflow rate of $\sim 600 M_\odot \yr^{-1}$, assuming a
covering fraction of $\delta \simeq 0.2$.
The cooling and ejection back to the ISM of large quantities of mass, make
the feedback process not only of energy (heating), but also of mass
(Pizzolato \& Soker 2005).
\newline
{(4)} Such SMW jets can inflate the `fat' bubbles
that are observed in many CFs, in clusters, groups of galaxies, and in elliptical galaxies
(Sternberg et al. 2007; Sternberg \& Soker 2008a, b).
{{{  The same holds for precessing jets (Sternberg \& Soker 2008a; Falceta-Goncalves et al. 2010)
    or a relative motion of the medium (Morsony et al. 2010). }}}
\newline
{(5)} The same mechanism that form SMW jets in CF clusters and galaxies,
can expel large quantities of mass during galaxy formation, and might
explain the SMBH-bulge mass correlation (Soker 2009; {{{  section 4.3 here). }}}

Following the problems with maintaining AGN feedback at galaxy formation by
the Bondi accretion (sec. \ref{sec:bondi}), and the discussion in section \ref{sec:cold},
I turn to elaborate on the possibility that the feedback at galaxy formation is similar
to that in CF clusters.

\subsection{A moderate cooling flow (CF) phase at galaxy formation}
\label{subsec:cooling2}

In their study of galaxy formation with AGN feedback, Croton et al. (2006)
consider the formation of an extended (up to the virial radius of the dark halo)
hot atmosphere, and that gas from the central region of this atmosphere
might be accreted onto a central object through a CF.
In this section I take upon this, and consider the moderate CF model
with a cold feedback.
Namely, the gas that feeds the SMBH originates in an extended region as cooling blobs,
and not as a Bondi type accretion. Heating of the inner regions by the AGN is
a significant process, and the AGN activity level is regulated by the accretion
of cold blobs from an extended region.

I differ from Croton et al. (2006) in a crucial manner.
In their model the main growth of the SMBH occurs in the `quasar mode' at high redshifts
($z > 2$).
During the CF phase of Croton et al. (2006) the accretion rate is low, and the
mass of the SMBH does not increase by much.
{{{  As shown in section 4.3, }}} in the present model, the correlation between the SMBH
mass and the bulge mass is set by a feedback process.
For the feedback process to be efficient, I require the presence of a CF.
Therefore, in the present model the CF appears much earlier than in the
model discussed by Croton et al. (2006).
{{{  The redshift at which the moderate CF operates to form stars and feed the SMBH
is determined by the specific model of galaxy growth that one uses.
I do not argue for a different global feeding of gas to the galaxy, and therefor the
SMBH growth period, as well as that of the bulge,
is as in the specific model, e.g., at $z>2$ in the model of Croton et al. (2006),
and in the redshift range of $z \sim 1-5$ in the model of Bower et al. (2006, 2008).
I limit myself to claim that a CF phase takes place during the main growth phase.
The proposed process can be incorporated in large scale simulations as sub-grid
physics. Namely, when mass is flowing to feed the central region, for both star formation
and SMBH growth, one can use the proposed mechanism to operate the feedback process
(as the inner region is not resolved by the large scale simulations).   }}}

I start when the SMBH mass is already $\sim 0.1-10\%$ of its final mass,
and the bulge/galaxy is already in the formation process.
However, most of the stars in the bulge (or elliptical galaxy) have yet to be formed,
and so does the SMBH mass.
Cold streams that feed the galaxy (Dekel et al. 2009a, b) are assumed to be
shocked at $r \ga 10 \kpc$, such that most of the mass resides in the hot phase.
The cooling time of the hot phase is short, and thermal instabilities lead
to the formation of cool blobs that fall inward.
{{{  These cool blobs are not much cooler than their surrounding for most of their
journey inward (Pizzolato \& Soker 2010).
Namely, they are not as cold as the gas discussed in section \ref{sec:cold}.
The cool blobs are not much denser than their hot surroundings, and they can
be expel relatively efficiently by the jets launched by the AGN when AGN activity is
high (see section 4.3). }}}
A fraction of these blobs feed the SMBH.
Heating by the SMBH activity, mainly by jets, facilitated the formation of
this structure.
The jets launched by the accreting SMBH not only heat the gas, but as is the case
in the cold feedback mechanism (Pizzolato \& Soker 2005), the jets accelerate large
quantities of gas outward. A structure similar to that in cluster CF at
low redshifts has been formed.
Table 1 compares the properties of the two types of the moderate CF models.
\begin{table}
\caption{Comparing Moderate Cooling Flow Models$^{(1)}$  }
\label{tab:data} \medskip
\small
\begin{tabular}{|l|c|c|}
\hline
Property$^{(2,3)}$         & Low $z$ Clusters           & Galaxy formation \\
\hline
\hline
Central $e^{-}$ density   $(n_{ce})$        &   $0.1 \cm^{-3}$           &  $10 \cm^{-3}$    \\
Central temperature       $(T_c)   $        &   $3 \times 10^7 \K$       &  $10^6 \K$       \\
System age                 $(\tau_{\rm age})$ &   $10^{10} \yr$            &  $10^9 \yr$      \\
Central cooling time$^{(4)}$ $(\tau_c)$       &   $5 \times 10^8 \yr $     &  $10^5 \yr $        \\
Cooling radius$^{(5)}$      $(r_c)$           &   $100 \kpc$               &  $30 \kpc$        \\
Dynamical time at               &              &    \\
$\qquad$ $0.01r_c$ $(\tau_{d1})$  &   $10^6 \yr$               &  $5\times 10^5 \yr$  \\
$\tau_c/\tau_{d1}$                          &      $500$                 &  $0.1$               \\
Raw cooling rate$^{(6)}$    $(\dot M_x)$      & $10-10^3 M_\odot \yr^{-1}$ &  $10-100 M_\odot \yr^{-1}$  \\
Star formation rate        $(\dot M_\ast)$   & $0-0.1 \dot M_x$           &  $\sim 0.1-0.5 \dot M_x$    \\
\hline
Source of gas feeding                       &  Cold blobs from an        & Cold blobs from an          \\
$\quad$ the SMBH                            &  $~$ extended region.      & $~$ extended region   \\
                                            &                           & +  inner supersonic inflow. \\
\hline
Fate of most gas expelled      & Inflating large       & Expelled from the galaxy.    \\
from the inner region          & low-density bubbles        &   \\
\hline
                               &  (a) Depart from 100\%          &  (a) Most of the ISM is  \\
                               &   efficiency in suppressing     & susceptible to SMBH jets.      \\
Results and implications       &   star formation.               & (b) Expelling huge amounts  \\
of the cold feedback           &  (b) Shallow entropy profile.   & of mass from the galaxy.      \\
 mechanism                      &  (c) Massive outflows that     &  (c) Can account for the    \\
                               &      can inflate 'fat bubbles'. &  $M_{\rm BH} - M_{\rm bulge}$ correlation  \\
                               &   (d) Can operate without               &   (Soker 2009; $\S 4.3$).    \\
                               &       the need for failed               & (d) Can operate without        \\
                               & Bondi accretion ($\S \ref{sec:bondi}$). & the need for failed  \\
                               &                                         & Bondi accretion.    \\
\hline
\end{tabular}
\caption{\footnotesize
(1) In the {\it Moderate CF model} heating is important, but cooling of gas to low
temperatures does occur, although at a much lower rate than that expected if no heating exists.
\newline
(2) The values are crude, and most are given to an order of magnitude.
\newline
(3) Some of the time scales are calculated by using equations from section \ref{sec:bondi}.
\newline
(4) The cooling time in galaxy formation is lower even, as a zero metallicity was assumed here,
but some metals will be present at an age of $\sim 10^9 \yr$.
\newline
(5) The cooling radius $r_c$ is the radius at which the radiative cooling time (no heating included)
equals the age of the system.
\newline
(6) Raw cooling rate is the mass cooling rate if no heating was presence.
}
\end{table}

There are, however, two prominent qualitative differences between the proposed CF model at
galaxy formation and that in CF in clusters of galaxies.
\newline
(1) In the case of galaxy formation the cooling time in the inner $\sim 1 \kpc$ is shorter than
the inflow time (section \ref{sec:bondi}).
The gas cools very rapidly, pressure support is lost, and
an inward supersonic flow is formed (Soker \& Sarazin 1988).
In clusters, on the other hand, the cooling time is longer than the dynamical time at all radii,
and no such flow is formed; the gas feeding the SMBH is cold, but it originates in the hot phase.
Some discussion of this feeding mode is given by Croton et al. (2006).
In addition to this supersonic flow, cold blobs formed from the hot phase
at larger radii feed the SMBH as well, as in the cold feedback model for heating CF
in clusters (Pizzolato \& Soker 2005, 2010; Soker 2006; 2008a).
\newline
(2) In clusters, the region where cooling takes place is $\sim 10-50 \kpc$, or about
a fraction $\sim 0.1-0.3$ of the cooling radius.
Most of the ICM mass resides outside this radius.
Any AGN activity can move mass from inner regions to regions further out,
but the huge amount of mass in the outer regions prevents
the mass to flow to very large distances from the center.
The situation is different in the proposed CF model at galaxy formation, where
most of the mass reside in regions having a cooling time shorter than the age
of the system.
The CF is a major process. Still, the heating by the central AGN ensures
the presence of a hot phase, and prolongs the time it takes the gas to cool.
That most of the mass is residing in the hot phase ensures an efficient coupling of
jets launched by the AGN to the ISM.
This coupling will be particularly efficient when the mass inflow rate is
$\ga 10^3$ times the accretion to the SMBH, a ratio that might
account for the SMBH$-$bulge masses correlation (Soker 2009; {{{  section 4.3 below). }}}

It is interesting to compare the model proposed here with that discussed by Binney (2004).
In both models heating, {{{   aided by reduction in radiative cooling rate (Cantalupo 2010), }}}
by the AGN activity is crucial, and in both models the ejection
of large quantities of gas out from the galaxy takes place.
The main difference is that Binney (2004) attributes all star formation to gas that
was never heated to about the virial temperature.
Instead, I suggest that a large fraction of the gas is shocked to the virial temperature,
and is further heated by the AGN to prolong its hot phase.
Most of the gas stays in the hot phase for a time longer than
the dynamical time, and forms a (pseudo) static medium.
Large quantities of this gas later cool, and form stars, as observed in cluster CF
(but at lower efficiency).
As the hot gas is much more susceptible to AGN activity, it allows for a feedback process from
the AGN to work, and determine the masses ratio of the SMBH and bulge (Soker 2009).

{{{
\subsection{Correlation of SMBH-bulge masses}
\label{sec:correlation}

In this section I briefly summarize the derivation of the correlation between the SMBH
mass and the host galaxy bulge mass (Soker 2009). I will not repeat all steps and
will not explain all assumptions, as they are in that paper. I will, however, make
some modifications to account for the present proposal of a CF phase at galaxy formation,
and to incorporate the new results of Soker \& Meiron (2010).

The basic assumptions are as follows.
\newline
(1) The feedback mechanism is driven by jets.
\newline
(2) The properties of jets launched by SMBH have some universal properties.
As is shown in a new paper (Soker \& Meiron 2010), the basic property is the
the ratio of the total momentum discharge in the jets to the quantity $\dot M_{\rm acc} c$:
$\epsilon_p \equiv \dot M_{f} v_f /(\dot M_{\rm acc} c)$,
where $\dot M_{\rm acc}$ is the accretion rate onto the SMBH,
$\dot M_{f}$ is the mass flow rate into the two jets, and $v_f$ is the jets' speed.
Soker \& Meiron (2010) find from statistical analysis of tens of galaxies that in the
feedback model for the correlation  $\epsilon_p = 0.038 \pm 0.06$.
\newline
(3) The mass flowing in at early stages, i.e., the mass available for star formation,
is very large. Namely, the mass that is converted to stars is limited by the feedback mechanism and
not by the mass available in the SMBH surroundings. This is supported by studies of galaxy formation
(e.g., Bower et al. 2008).
\newline
(4) There is a relative transverse (not to be confused with the radial inward velocity)
motion between the SMBH and the inflowing mass of $v_{\rm rel} \simeq \sigma$.
\newline
(5) The cooling surrounding mass $M_s$ that resides at a typical distance $r_s$ and
having a density $\rho_s$ (see below), is
flowing inward at a velocity of $\sim \sigma$.
Thus, $\dot M_s \simeq 4 \pi r_s^2 \sigma \rho_s$,
and it is resupplied on a time scale of $\sim r_s/\sigma$.
This mass will mainly form stars if it is not expelled by the jets.
This assumption is in the heart of the proposed CF phase at galaxy formation.
Namely, that the inflowing gas, be it a cooling flow that reaches a fast speed
at $r \sim 1 \kpc$ (where the cooling time is short) or composed of cool clumps,
is vulnerable to the jets launched by the central SMBH. The inflowing gas or clumps, are cooler than
the virial temperature but not by much, hence their density is not much lower than
that of the surrounding hot gas.
They are not the very dense clumps at low temperatures that were discussed in
section \ref{sec:cold}, and they can be expel by the jets.

We note that $\dot M_s \gg \dot M_{\rm acc}$, as only a small fraction of the inflowing gas at
scales of $\sim 0.1-10 \kpc$ is accreted by the SMBH.

If the jets penetrate through the surrounding gas they will be collimated by that gas,
and two narrow collimated fast jets will be formed, similar to the flow structure in
the simulations of Sutherland \& Bicknell (2007).
By fast it is understood that the jet's velocity is not much below its original velocity.
If, on the other hand, the jets cannot penetrate the surrounding gas they will
accelerate the surrounding gas and form SMW (slow-massive-wide) outflow (Soker 2008).

I now derive (Soker 2009) the conditions for the jets not to penetrate the surrounding
gas,  but rather form a SMW outflow.
Let the jets from the inner disk zone have a mass outflow rate in both directions
of $\dot M_f$, a velocity $v_f$, and let the two jets cover a solid angle of
$4 \pi \delta$ (on both sides of the disk together).
The density of the outflow at radius $r$ is
\begin{equation}
\rho_f =  \frac {\dot M_f}{4 \pi \delta r^2 v_f}.
\label{eq:rhof}
\end{equation}
Let the jets encounter the surrounding gas residing within a distance
$r_s$ and having a typical density $\rho_s$; this is the inflowing cooling gas,
that if it is not expelled will form stars.
The head of each jet proceeds at a speed $v_h$ given by the balance
of pressures on its two sides.
Assuming supersonic motion this equality reads
$\rho_s v_h^2 = \rho_f (v_f-v_h)^2$, which can be solved for $v_h$
\begin{eqnarray}
\frac {v_f}{v_h}-1 =
\left( \frac {4 \pi \delta r_s^2 v_f \rho_s}{\dot M_f} \right)^{1/2}
\simeq \left( \frac {\delta \dot M_s v_f }{\dot M_f \sigma} \right)^{1/2}
\nonumber \\
=1225
\left( \frac {\dot M_s/\dot M_f}{10^4} \right)^{1/2}
\left( \frac {\delta}{0.1} \right)^{1/2}
\left( \frac {v_f}{c} \right)^{1/2}
\left( \frac {\sigma}{200 \km \s^{-1}} \right)^{-1/2}.
\label{eq:vh1}
\end{eqnarray}
where in the second equality the mass inflow rate $\dot M_s \simeq 4 \pi \rho_s \sigma r_s^2$
(by assumption 5), has been substituted.
The time required for the jets to cross the surrounding gas and break out of it is given by
\begin{equation}
t_p \simeq \frac {r_s}{v_h} \simeq
\frac{r_s}{v_f}
\left( \frac {\delta \dot M_s v_f }{\dot M_f \sigma} \right)^{1/2}
=4 \times 10^6
\left( \frac {r_s}{1 \kpc} \right) \yr ,
\label{eq:tp1}
\end{equation}
where in the last equality the same values as in equation (\ref{eq:vh1}) have been used.

If there are no changes in the relative geometry of the SMBH and inflowing mass,
the jets will rapidly penetrate the surrounding gas and expand to large distances.
In this case the jets will not deposit their energy in the inflowing gas.
For an efficient deposition of energy to the inflowing gas, we require that there will
be a relative transverse (azimuthal) motion between the SMBH and the inflowing gas,
such that the jets continuously encounter fresh mass.
The relevant time is the time that the transverse motion of the jet
crosses it width $\tau_s \equiv D_j/v_{\rm rel} \simeq D_j/\sigma$,
as by our assumption 4 the relative velocity is $v_{\rm rel} \simeq \sigma$.
The width of the jet at a distance $r_s$ from its source is $D_j=2 r_s \sin \alpha$,
where $\alpha$ is the half opening angle of the jet. For a narrow
jet  $\sin \alpha \simeq \alpha \simeq (2 \delta)^{1/2}$, and
\begin{equation}
\tau_s = \frac {2 (2 \delta)^{1/2} r_s}{v_{\rm rel}} =4.4 \times 10^6
\left( \frac {r_s}{1 \kpc} \right)
\left( \frac {v_{\rm rel}}{200 \km \s^{-1}} \right)^{-1}
\left( \frac {\delta}{0.1} \right)
\yr .
\label{eq:taus}
\end{equation}

The demand for efficient energy deposition, $\tau_s \la t_p$, reads then
\begin{equation}
\frac {\dot M_s}{\dot M_f} \ga
8   \frac {v_f \sigma }{v^2_{\rm rel}}.
\label{eq:ms1}
\end{equation}
This result can be understood as follows. The ratio ${v_f \sigma }/v^2_{\rm rel}$ comes from the ratio
of the ram pressure of the narrow jet to that of the ambient gas which disturbs the jet, and from
the relative transverse motion of the jet and the abient gas.
The number 8 comes from the geometry of a narrow jet with a relative transverse velocity to
that of the ambient gas.
Using the definition $\epsilon_p \equiv \dot M_{f} v_f /(\dot M_{\rm acc} c)$ from assumption 2, we derive
\begin{equation}
\frac {\dot M_s}{\dot M_{\rm acc}} \ga
8 \epsilon_p \frac {\sigma c}{v^2_{\rm rel}}
= 480
\left( \frac {\epsilon_p}{0.04} \right)
\left( \frac {\sigma}{200 \km \s^{-1}} \right)^{-1}
\left( \frac {\sigma}{v_{\rm rel}} \right)^2.
\label{eq:mscluster}
\end{equation}
Again, it is expected that in its formation phase the galaxy will not be fully relaxed,
and that the relative transverse velocity of the AGN and the inflowing gas will be of the
order of the stellar dispersion velocity, i.e., $v_{\rm rel} \simeq \sigma$

The accretion rate $\dot M_{\rm acc}$ is the accretion rate onto the SMBH, and the
inflow rate of the surrounding gas is assumed to form stars in the bulge (if it is not
expelled by the jets).
If the inflow rate is above the value given by equation (\ref{eq:mscluster}), the
deposition of energy by the jets is efficient enough to expel the mass back to large
distances and heat it (Soker 2008b).
The interaction of the (narrow or wide) jets blown by the SMBH with the inflowing gas will
form a wide outflow (SMW jets), that will expel more of the hot gas that is vulnerable to
the jets.
Namely, the jets blown by the SMBH will not allow the bulge to form stars at a rate
larger than the value of $\dot M_s$ given by equation (\ref{eq:mscluster}).
Following Soker (2009) then, the SMBH to bulge mass ratio is about equal to
$\dot M_{\rm acc}/{\dot M_s}$.
Equation (\ref{eq:mscluster}) yields
\begin{equation}
M_{\rm BH} \simeq 0.002 M_{\rm bulge}
\left( \frac {\epsilon_p}{0.04} \right)^{-1}
\left( \frac {\sigma}{200 \km \s^{-1}} \right)
\left( \frac {\sigma}{v_{\rm rel}} \right)^{-2}.
\label{eq:mBH2}
\end{equation}
The last equation closes the feedback cycle, in showing that a correlation can be driven
by jets blown by the SMBH into the hot ISM.
A key issue is that the medium is in the hot phase such that its density is not too high,
and therefore it is vulnerable to the action of the jets. This hot phase feeds the SMBH via
the process of a moderate cooling flow.
}}}

\section{SUMMARY}
\label{sec:summary}

Results from recent years show that the process of galaxy formation requires
non-gravitational energy source not only to heat the gas, but also to expel
large quantities of gas out from the galaxy  (Bower et al. 2008).
To efficiently eject the ISM from the galaxy by AGN activity the gas must be in the hot
phase, namely, its temperature must be about the virial temperature (e.g., Hopkins \& Elvis 2009).
The conclusion from these studies is that most of the ISM during galaxy formation
must evolves through the hot phase. This gas has a short cooling time, and a
cooling flow (CF) is formed in the still-forming galaxy.

In the present paper I assumed that the $M_{\rm BH}-M_{\rm bulge}$ correlation is
determined by an AGN feedback mechanism that operates during a cooling flow phase at
galaxy formation {{{  (section 4.3). }}}
It is determined by the sense that the AGN activity limits the ISM mass that is eventually
converted to stars by expelling it out of the galaxy.
I showed that the Bondi accretion cannot operate during that phase (section \ref{sec:bondi}),
and discussed the requirement that the ISM be in the hot phase (section \ref{sec:cold}).
As the radiative cooling time of the hot phase is relatively short (eq. \ref{eq:time1}),
to maintain a hot phase the infalling gas should be shocked at a radius of $R \ga 5 \kpc$.
Heating by the AGN facilitates the formation of the hot phase.
As with CFs in clusters of galaxy, the heating and ejection of the ISM is done
by jets, rather than by radiation.

The short radiative cooling time implies the formation of a CF in the the inner
region of the newly formed galaxy, but one that is substantially heated by the AGN activity.
As Bondi accretion fails, the feeding of the SMBH is done via two channels.
Like in the \emph{cold feedback mechanism} in clusters of galaxies
(Pizzolato \& Soker 2005, 2010; Soker 2006, 2008b), cold blobs are falling from an extended region.
The second channel is a cold supersonic inflow in the inner $\la 1 \kpc$ of the galaxy,
where cooling time is shorter than the inflow time; such an inflow was studied by
Soker \& Sarazin (1988).
The AGN jets can be efficient enough to expel a large fraction of the inflowing gas and the
gas in the hot phase (Soker 2009) out of the galaxy. Some of the cooling gas will form stars and
feed the SMBH.
A cooling flow where a large fraction of the cooling gas is expelled form the inner region
(and some fraction forms stars, and a small fraction is accreted by the SMBH)
 is termed a \emph{moderate CF model} (Soker et al. 2001, Soker \& David 2003).
In section \ref{subsec:cooling2} the moderate CF model in clusters of galaxies and at galaxy
formation are compared, and summarized in Table 1.

The moderate CF model proposed here at galaxy formation can be applicable to the sample
of obscured AGN studied by Brusa et al. (2009).
The median value of $L_{\rm AGN}/L_{\rm Edd}$ in their sample is $\sim 2-10 \%$.
This gives a SBMH growth time of $\sim 10^9 \yr$, similar to the growth time of the stellar
population due to star formation.
I propose that a moderate CF exists in these galaxies during the high star
formation rate.

\acknowledgements
{{{  I thank an anonymous referee for helpful comments. }}}
This research was supported by the Asher Fund for Space
Research at the Technion, and the Israel Science foundation.

\end{document}